\documentstyle[preprint,aps]{revtex}
\begin{document}
\draft
\title{
Relevance of Interchain Hopping in Correlated Hubbard Chains
}
\author{
S. P. Strong$^{(1)}$}

\address{
$^{(1)}$NEC Research Institute,
4 Independence Way, Princeton, NJ 08540
}
\date{\today}
\maketitle
\begin{abstract}

We demonstrate that the hopping of singlet electron
pairs between Hubbard chains is
relevant in the renormalization group sense
if appropriate correlations exist between the
chains.
\end{abstract}

\narrowtext

Since the discovery of superconductivity in the
copper oxide systems by Bednorz and M\"uller
\cite{BM}
various mechanisms have been proposed to account
for the high superconducting transition temperatures
in these materials.  One candidate,
the interlayer pair hopping
mechanism \cite{AWH},
offers an explanation for the high transition
temperatures in terms of a ``blocked'' single particle
interlayer hopping and a relevant
pair hopping between layers.
An explanation of how such blocking
of the single particle hopping
can occur in a strongly correlated system,
in spite of the  {\it relevance}
 of the  hopping operator,
has been given elsewhere \cite{incoherence},
as has an argument that this effect
has already been experimentally observed \cite{magic}.
Using the assumption of momentum space
diagonality
\cite{kdiag},
pair hopping operators have been
constructed which are relevant and
lead to mean-field gap equation with a high superconducting
transition temperature.
In contrast, this paper will instead discuss the circumstances
under which the real-space-local pair
hopping operator
 $O_s(j) =
\left( \psi^+_{1,\uparrow}(j) \psi^+_{1,\downarrow}(j+1)
- \psi^+_{1,\downarrow}(j) \psi^+_{1,\uparrow}(j+1) \right)
\left(
 \psi_{2,\uparrow}(j) \psi_{2,\downarrow}(j+1)-
 \psi_{2,\downarrow}(j) \psi_{2,\uparrow}(j+1) \right)~+~h.c.$,
can become relevant for a model of coupled
Hubbard chains, the prototypical model
of an anisotropic, strongly correlated
system.  How a generalization to
Hubbard planes might work will also
be discussed.

The motivation for this work came from the
recent
discovery that
the one dimensional Heisenberg model
has a peculiar form of off-diagonal long
range order \cite{pwa_conj,singlet}
similar to that of the Fractional
Quantum Hall Effect \cite{rgm}.  In particular,
the overlap between the groundstate of the
Heisenberg model with $N$ sites and the
state obtained by adding a pair of sites
in a singlet configuration to the groundstate
of the $N-2$ site Heisenberg model is finite.  This
order has important implications for the
renormalization group
relevance of the operator, $O_s$,
which hops singlet pairs of electrons between
pairs of Hubbard chains as we now discuss.

The renormalization group relevance or irrelevance
of the pair hopping operator,
is determined by
its scaling dimension.
This can be obtained by examining the exponent with
which its correlation functions decay for large
separations.
If $\langle O^+_s(j) O_s(0) \rangle$
decays at long distances like $j^{-2d}$,
then $d$ is the scaling dimension
of $O_s$. If
the scaling
dimension is less than $2$,
then
$O_s$ is relevant in the infrared.  In
perturbing about the independent chains case,
the calculation can be simplified since
the $O_s$ correlation function factorizes into the
product of the singlet pair correlation functions
in the isolated chains.  The relevance or
irrelevance of $O_s$ is then determined by the
divergence or failure to diverge of the
isolated-chain, pair susceptibility. For
all positive $U$, the pair susceptibility is
finite and $O_s$ is irrelevant for Hubbard
chains coupled only by singlet pair hopping. We will
see that this is not necessarily the case for
Hubbard chains which are correlated.

To see how this comes about and what sort of
correlations are required,
we first discuss
the large $U$ Hubbard model, where the
exact wavefunction of Ogata and Shiba
\cite{OgataShiba} can be used, together
with the results of \cite{singlet}.
The Ogata-Shiba wavefunction is
a product of spin and charge wavefunctions. The
charge wavefunction is a spinless fermion determinant
with the positions of the spinless fermions given
by the real electron positions; the spin wavefunction
is (for appropriate boundary conditions) the groundstate
wavefunction of the Heisenberg model given by
Bethe's solution \cite{bethe} with the positions of the
electrons identified with the sites of the
Heisenberg chain.
The insertion of a singlet pair of electrons into
a large $U$ Hubbard model acts on this wavefunction
in a straightforward way, inserting a pair of
spinless fermions into the charge wavefunction and
a  singlet pair of spins into
the Heisenberg chain.  The result of \cite{singlet}
is that the spin state obtained in this way has
finite overlap with the true groundstate of the
Heisenberg model with the appropriate number of
sites.  This allows us to write part of the action on the
groundstate wavefunction of a singlet insertion
purely in terms of the charge wavefunction:
$
 \psi^+_{el,\uparrow}(j) \psi^+_{el,\downarrow}(j+1)
\rightarrow ~overlap~  \psi^+_{sf}(j)\psi^+_{sf}(j+1)$
 The overlap is just a constant, except for the
following consideration: since the momenta
of the groundstates of the $N$ and $N+2$
site Heisenberg models differ by $\pi$, the finite
overlap changes sign every time the inserted singlet is
displaced one site in the Heisenberg model
or past one electron in the large
$U$ Hubbard model.
This means that
$ \psi^+_{el,\uparrow}(j) \psi^+_{el,\downarrow}(j+1)
\rightarrow ~const~ e^{i \pi \sum_{l<j} n_j}
 \psi^+_{sf}(j)\psi^+_{sf}(j+1)$
where the constant is a number of order unity (see \cite{singlet}).
The correlation function of
$\psi^+_{sf}(j)\psi^+_{sf}(j+1)$
with
its Hermitean conjugate is just a
free fermion correlation function and so decays like
$j^{-2}$. This would make pair hopping marginally
relevant even at infinite $U$, except for the
presence of the alternating sign
induced by $e^{i \pi \sum_{l<j} n_j}$.
To make the pair susceptibility of an
isolated Hubbard chain diverge we could
modify its correlations so
that there was an expectation value for
$-1$ raised to a power given by the number of
particles to the left of site $j$.  This
would involve radically changing the
correlation structure of the chain,
in effect requiring electrons to come in
pairs.
The only
obvious way to do this is to change the sign of
$U$, in which case finding a diverging pair
susceptibility is not surprising.  Negative
$U$ models are unlikely to be relevant to
the cuprates. Fortunately we
are interested in the pair susceptibility only
for what it can tell us about the relevance of
the interchain pair hopping operator. In that
case, the operator that needs to acquire and
expectation value is
$-1$ raised to the power given by the total
number of
particles in {\it both} chains to the left
of site $j$. States with this
property can be constructed for
positive $U$ and a reasonable choice for
the interchain coupling as explained below.

In order to discuss which types of
correlations between
the chains satisfy this requirement it
is convenient to introduce the Abelian bosonization
formalism or Luttinger liquid approach
to the one dimensional Hubbard model \cite{Haldane,1dHubb}.
In that language the effective low energy Hamiltonian
of the Hubbard model becomes:
\begin{equation}
H = \frac{1}{4 \pi} \int dx~ \left(
v_{\rho} K_{\rho} (\partial \Theta_{\rho})^2 +
v_{\rho} K_{\rho}^{-1} (\partial \Phi_{\rho})^2 +
v_{\sigma} K_{\sigma}  (\partial \Theta_{\sigma})^2 +
v_{\sigma} K_{\sigma}  (\partial \Phi_{\sigma})^2
\right)
\label{eq:ham_hubb}
\end{equation}
where $K_{\rho}$ is $U$ dependent ranging from
$1$ at $U=0$ to $\frac{1}{2}$ as $U \rightarrow \infty$,
$K_{\sigma}$ is fixed at unity by $SU(2)$ invariance and
bosonic representations of all operators can be
constructed from those of the electron operator \cite{Haldane}:
\begin{equation}
\Psi^{\dagger}_{\uparrow}(x) \sim
\sqrt{ \frac{\partial \Phi_{\uparrow}(x)}{ \pi}}
\sum_{m ~odd} \exp\left(i [m \Phi_{\uparrow}(x) +
 \Theta_{\uparrow}(x)]\right)
\label{eq:elop}
\end{equation}
In particular,
the bosonized form of the pair insertion operator
is given by:
\begin{eqnarray}
\Psi^{\dagger}_{\uparrow}(x)
\Psi^{\dagger}_{\downarrow}(x) & \sim &
  \frac{1}{ \pi}
 \sqrt{ \partial \Phi_{\uparrow}(x) \partial \Phi_{\downarrow}(x)}
\sum_{m,n ~odd} \exp\left(i [m \Phi_{\downarrow}(x) +
 \Theta_{\uparrow}(x)
 + n \Phi_{\uparrow}(x) +
 \Theta_{\downarrow}(x)]\right) \nonumber \\
 & \sim & \rho_0 \exp\left(i \Theta_{\rho}(x)\right)
 \cos \Phi_{\sigma}(x) + \dots
\end{eqnarray}
where $\rho$ and $\sigma$ subscripts
label the symmetric and antisymmetric
combinations of the up and down spin phase fields.
The bosonized form of the pair hopping operator
is given by:
\begin{equation}
O_s(x) \sim  \rho_0^2 \cos \Theta_{\rho,A}(x)
[\cos \Phi_{\sigma,A}  + \cos \Phi_{\sigma,A}  ] + \dots
\label{eq:bospairhop}
\end{equation}
where $S$ and $A$ subscripts label symmetric and antisymmetric
combinations of the phase fields of the two chains.
The leading contribution to
the bosonized form of $\exp(i \pi \sum_{l<j} n_{1,\uparrow}(l) )$
is just $exp\left( i \Phi_{\uparrow}(ja) \right) + ~h.c.$
so that the sign factor relevant to coupled chains,
represented by the operator
$\exp(i \pi \sum_{\sigma,a} \sum_{l<j} n_{a,\sigma}(l) )$,
will acquire an expectation value if any of
the operators:
\begin{eqnarray}
O_1 & = &
 \cos \Phi_{\rho,S} \\
O_2 & = &
 \cos \Phi_{\rho,A} \\
O_3 & = &
 \cos \Phi_{\sigma,S} \\
O_4 & = &
 \cos \Phi_{\sigma,A}
\end{eqnarray}
acquire expectation values
(provided things are not contrived so that
several of the above acquire expectation
values which sum to zero).
The first of the cosines can not acquire a
non-zero expectation value
for a generic filling fraction while,
following equation
\ref{eq:bospairhop},
 the
second  would introduce a gap into the antisymmetric,
 charge sector
of the model so that the correlation functions
of $O_s$
would decay exponentially.
However, an expectation
value for either of the other two cosine operators
will dramatically enhance the correlations
of $O_s$, as
can clearly be seen from equation \ref{eq:bospairhop}
as well as from the more physically intuitive
argument we have given for large $U$.

It is clear that to produce an expectation
value for a cosine operator whose argument is
a combination of the spin phase fields in the
different chains, we need to correlate the
spin structure between the two chains.
Phenomenologically,
many of the cuprate compounds exhibit ``spin gap'' behavior \cite{spingap}
and the fact that this behavior is clearest for the
two-layer per unit cell compounds
\cite{andyspingap,nospingap}
suggests that
the spins in different layers are indeed correlating
at low temperature in the cuprates.
Further, density matrix renormalization group
studies of Hubbard chains coupled by an interlayer hopping
find a spin gap phase for a broad range of
hopping strengths \cite{white}.
It  is therefore natural to seek two-chain
spin sector states which posses a spin gap and an expectation
value for either $ \cos \Phi_{\sigma,S} $
or $ \cos \Phi_{\sigma,A}$. One candidate state
is the analogue of the disordered state found
for coupled Heisenberg chains \cite{meandy}
where $\cos \Theta_{\sigma,A}$ and $  \cos \Phi_{\sigma,S}$
take on expectation values.  This state should
result if the dominant coupling between the
two chains is an antiferromagnetic
coupling of the small momentum pieces of the
spin density operators.  The spin coupling
 operator then
takes the form $ \cos \Phi_{\sigma,S} \cos \Theta_{\sigma,A}$
and is marginally relevant for an antiferromagnetic
coupling.  Since the symmetric and
antisymmetric sectors are decoupled
this should lead to expectation values for
 $\cos \Phi_{\sigma,S}$ and $\cos \Theta_{\sigma,A}$,
separately.  The dual state where
$ \cos \Phi_{\sigma,A}$
and $ \cos \Theta_{\sigma,S}$ acquire expectation values
also has the correct properties, but
we are unaware of any operator which
would naturally give rise to this state.

In the former state, as a consequence of the expectation
value for $\cos \Phi_{\sigma,S}$,
the singlet
pair hopping, $O_s$,
contains a piece given by
$const. ~ \cos \Theta_{\rho,A}$.
This operator is relevant with scaling dimension
$K_{\rho}^{-1}$, ranging between
$1$ and $2$ for the positive $U$
Hubbard model \cite{renorm}, and can
be expected to
dominate the low energy physics.
In the infinite $U$ limit,
the charge sector of the model can
be re-fermionized, with the
operator, $\cos \Phi_{\sigma,S}$
becoming nothing more than
pair hopping of the
otherwise free, spinless fermions.
The equivalent, refermionized model is
two chains of
free spinless fermions
coupled only by a hopping between the chains of pairs
of fermions at opposite Fermi points.
In effect, hopping a singlet pair of electrons
is equivalent to hopping a pair of free holons.
(Note that the charge part of the
electron operator in one dimension is a semion,
not a spinless fermion, and that it is only the hopping
of a pair of electrons which can be
written in terms of the hopping of a
pair of spinless fermions, and then only in the large $U$ limit
for correlated chains;
the ``holon'' we are discussing is a true spinless fermion
and not equivalent to some of the other uses of the term.)
Although the one-dimensional,
two chain version can not acquire long range
pairing order of its holons,
a higher dimensional analogue of this state
is ideal for realizing the interlayer pair hopping
mechanism for high $T_c$ and one would expect the
hopping to act as an effective attractive interaction,
leading to holon pairing.

It is interesting to see how the invariance of
the pair insertion to a spatial interchange of
the electrons in a Cooper pair would occur in the above
framework.  The electron insertion is
equivalent to a singlet insertion plus a
holon pair insertion, and the holon pair insertion
is odd under a spatial interchange so that
the eveness of the electrons under spatial
interchange occurs only because the singlet insertion
is also odd under spatial interchange.
The holon and spinons are effectively $p$-wave
paired so that electrons are either $s$ or $d$-wave
paired, which are the same in one dimension.
If the correct two dimensional generalization
of the solution of the one dimensional Hubbard model
has an analogous version of spin-charge
separation to the one dimensional solution
then one expects to find the pairing of
 spinless fermion holons at
large $U$ in a
$p$-wave  channel, and,
since the singlet insertion
is odd under a $180^{\circ}$ rotation of the
singlet pair, the natural symmetry for the
electron pairing would actually be $d$-wave.
Since the singlet is tied to the real space lattice,
$d_{x^2-y^2}$ is the natural choice.

The resulting phenomenology is in excellent
agreement with that observed in the
bilayer cuprates, including the lack of
correlation of $T_c$ with $\rho_{ab}(T)/T$
emphasized by Anderson \cite{pers_com},
the strong correlation of
$T_c$ with the number of layers \cite{AWH},
the gap symmetry and
the presence of a spin gap phase.
 For materials with
one or three layers per unit cell
a theory of two coupled
chains is clearly  a bad starting point.
The evidence for the absence of a spin gap
in these materials
\cite{nospingap,andyspingap} is therefore
not inconsistent with the idea of
interlayer
spin correlations, which we require here
to change the
relevance of the pair hopping. The spins in
odd layer per unit cell compounds may
still correlate with each other, but, with
an odd number of layers per unit cell, there is no
natural way to put everything into singlets
and gap the entire spin spectrum.

Coupled Hubbard chains
have been studied previously in references \cite{t_rel},
\cite{khev} and
\cite{nagaosa}, none of which
considered the scenario envisioned
here for reasons which we now discuss briefly.  The
first set of works focused on the
relevance of single particle hopping
between Hubbard chains.
A single particle hopping operator
with an appreciable coefficient
should be present
in the cuprates;
we have chosen to ignore
the effects of single particle hopping
for several reasons.
First, the theoretical work of
\cite{incoherence} suggests that
for strongly correlated, anisotropic
systems the coherent effects of
single particle
hopping may
vanish, despite the relevance of
the hopping operator, while \cite{magic}
 argued that this effect has already been
observed experimentally.
Second,
photoemission data of
\cite{photem},
which have  an energy resolution of
8 $meV$,
show no evidence of the Fermi surface splitting
for the coupled planes in BISCO 2212,
despite that fact that
a coherent single particle hopping should
produce such a splitting for the
Fermi surfaces of the symmetric and
antisymmetric fermions and the scale
of the splitting predicted
by band theory calculations is of order tenths
of an electron volt \cite{phils}.
Lastly, the highly anisotropic
resistive
properties of the cuprates essentially
rule out coherent three dimensional
transport \cite{Ong} suggesting that
single particle interlayer hopping is not
dominating the low energy  physics
as would naively be expected.
We therefore believe it is appropriate to
consider a model where the single
particle hopping is neglected.

References \cite{khev} and
\cite{nagaosa}
focused on Hubbard chains
with an antiferromagnetic coupling
of
the $2k_F$ pieces of the spin
operators in the two chains.
These works were primarily concerned with
how an isolated two dimensional
Hubbard model
might be better understood through
the study of coupled Hubbard chains.
In this context those works noted
that the resulting expectation
values for
$\cos \Phi_{\rho,A}$, $ \cos \Theta_{\sigma,A}$
and $\cos \Phi_{\sigma,S}$ lead to
an enhanced tendency towards interchain
pairing and the formation of a spin-gap.
If one believes that high temperature
superconductivity and spin gap behavior
are present for isolated
copper oxide planes then these are
significant results.
On the other hand,
considered as a one-dimensional version
of coupled planes,
their state would be unfavorable to
interlayer
pair hopping
because the antisymmetric charge sector
has a gap, leading to  the exponential
decay of the correlations of $O_s$
and so is not a candidate for the effects
we are interested in.

Their state
is clearly  relevant to
Hubbard chains with a weak antiferromagnetic
coupling, since
the
coupling of the $2k_F$ pieces of the spin
operator is more relevant that the
coupling of the small momentum pieces.
On the other hand
there are a number of reasons
to prefer an analogue of the
state without a charge gap
for the cuprates.
One reasons is that, at half-filling, the
two dimensional analog of the
$2k_F$ piece of the spin operator in one dimension
is the Neel order
parameter. Coupling
the Neel order in two planes will
produce optical and acoustic magnons
rather than
a spin gap, so if one
attributes spin gap behavior to
interlayer coupling (rather than
intralayer  physics as intended in \cite{khev,nagaosa})
the coupling focused on in those works
does not lead to a spin gap in two dimensions.
Away from half-filling, there in no clear analogue of
the $2k_F$ piece of the spin operator and the
the spin operator should involve low energy
pieces with (if there is Fermi surface)
all momenta which span the Fermi surface. This
should disfavor a
state based on a particular wavevector, which would
be the analogue of $2k_F$.
Another factor favoring the
operator we consider is that the spin energy
 scale is much smaller
than the charge energy scale for the
large $U$ Hubbard model, so an operator
which seeks to modify only spin degrees of
freedom should be correspondingly more effective
at a given strength than one that require
the modification of both spin and charge degrees
of freedom.
Further, coupled plane states analogous to that of
\cite{khev,nagaosa} would have
an interlayer pairing
similar to that discussed in
\cite{Lee}.
The drawback  of such a state
is that an interlayer pair should transform
trivially under rotations in the $ab$ plane
\cite{Lee} whereas current experimental evidence
strongly supports a gap transforming with
$d$-wave symmetry \cite{squids}.
Finally,
there is the issue of the
exact effects of the
relevant single particle hopping operator
which should be present in any model
meant to be relevant to the cuprates.
Both the interchain correlations discussed here and
those discussed in \cite{khev} and \cite{nagaosa}
require that it be possible to effectively eliminate
$t_{\perp}$, the interchain hopping.  This can be done
as in \cite{nagaosa} for a more general model where the
spin-spin coupling between the chains in introduced
independent of the interchain hopping so that it
can dominate over $t_{\perp}$ simply because of its
coefficient.  However, if one takes the perspective that
any spin couping between the chains should arise
as a superexchange coupling obtained from
integrating out the
single particle hopping,
then the problem arises that $t_{\perp}$ is
more relevant than the spin-spin coupling and should dominate
the physics.
It is possible
 \cite{possible} that this operator is removed by incoherence
effects resulting from strong interactions
so that
it one can effectively integrate
out $t_{\perp}$ and consider the operators
generated at higher order, however,
no rigorous prescription exists for doing so since
it would require a full understanding of the
strong-coupling, incoherent fixed point proposed in
\cite{incoherence} which is not presently available.
In the absence of such an understanding, it is not
clear that
the operator which is more relevant at the
$t_{\perp} = 0$ fixed point need dominate
the flow away from the strong-coupling, incoherent fixed point.
We have therefore chosen to consider a
state dominated by the coupling of the non-alternating
pieces of the spin operator.

In summary, we have considered a model of coupled Hubbbard chains
different than those considered previously and have found that
appropriate correlations between the chains can lead to the
renormalization group relevance of the interchain pair
hopping operator.    The relevance can be considered to
result from the
hidden ODLRO of the one dimensional Heisenberg chain
\cite{singlet} together with interchain correlations.
The most reasonable candidate for such correlation
is a state resulting from coupling the small momentum
pieces of the spin operators in the two chains,
leading to a spin gap. The resulting state would
realize an interchain version of the interlayer tunneling
theory of superconductivity \cite{AWH}, provided that
the divergent, one-dimensional fluctuations
of a single pair of chains were stabilized by
some coupling to the other pairs in a three dimensionsal
array of such chains.  The natural generalization to
coupled planes should exhibit $d_{x^2-y^2}$ symmetry,
a spin gap and a high transition temperature dependent on
the interlayer coupling, but not on the in-layer
resistivity, in precise agreement with the phenomenology
of the high $T_c$ cuprates.

I gratefully acknowledge
helpful discussions with
P.~W.~Anderson and
J.~C.~Talstra
as well as financial support from
the NEC corporation.


%
%

%


\begin{references}
\bibitem{renorm} $K_{\rho}$ will
 be subject to some
renormalization from the interchain
couplings already intergrated out,
but,
since we are assuming that the
dominant operator involves only
spin phase fields and since the self-dual
nature of the first order $t_{\perp}$
prevents it from renormalizing $K_{\rho}$,
any such renormalization should be weak.
\bibitem{possible}
Integrating out an operator is only justifiable
if it acts incoherently or is irrelevant.
To see this consider the fact that,
for it to be possible to integrate an
operator out at all, its  spectral function must vanish
as $\omega \rightarrow 0$, otherwise the
some of the coefficients of the operators
generated at second order in perturbation theory
will in general be infinite.  As an example, consider the
fact that the identity operator is
generated with a coefficient which gives the contribution
to the ground state energy at second order in perturbation theory:
\begin{equation}
\Delta E = - \sum_m \frac{| \langle 0 | H^{'}| M \rangle |^2}{E_m}
\end{equation}
or if $H^{'} = \lambda O$ then
\begin{equation}
\Delta E =  - \lambda^2 \int~\frac{d \omega}{\omega} \rho_O(\omega)
\end{equation}
where $\rho_O(\omega)$ is the spectral function for the operator
$O$ defined to be
$\sum_m~| \langle 0 | O | M \rangle |^2~\delta(E_M-\omega)$.
For spectral functions which do not vanish
as $\omega \rightarrow 0$
the shift is logarithmically divergent, signalling
that
the perturbation,
$H^{'}$, must be treated in nearly degenerate perturbation
theory, not in ordinary perturbation theory and the
usual approach to integrating it will generate nonsensical
answers.
Our definition for coherence is essentially
that the spectral function
does not vanish sufficiently
quickly.
For $t_{\perp}$ this situation is complicated
by the effects of the Fermi statistics of the electrons
on the spectral function.  For free particles, the
spectral function of $t_{\perp}$ vanishes identically
since Fermi statistics prevent any hops in the ground state,
however, $t_{\perp}$ clearly needs to be treated in degenerate
perturbation theory and is coherent.  We believe that the essential
point still holds that the coefficients of
operators generated at second order
will diverge unless the response to the perturbation is incoherent.
More detailed arguments will be given elsewhere [29].
\end{references}
\end{document}